\documentstyle[preprint,aps]{revtex}
\begin{document}
\bibliographystyle{unsrt}
\title{Isospin Breaking in the Pion-Nucleon Coupling \\
 from QCD Sum Rules}
\author{T.Meissner}
\address{Department of Physics, \\
Carnegie-Mellon University,
Pittsburgh, PA 15213, USA  \\}
\author{E.M.Henley}
\address{Department of Physics,University of Washington,\\ 
Box 351560, Seattle, WA 98195, USA \\}
\date{\today}
\maketitle
\begin{abstract}
We use QCD sum rules for the three point function 
of a pseudoscalar and two nucleonic currents in order
to estimate the charge dependence of the pion nucleon coupling
constant $g_{NN\pi}$ coming from isospin violation in the strong
interaction. The effect can be attributed primarily to the difference 
of the quark condensates $<{\bar u}u>$ and $<{\bar d}d>$.
For the splitting 
$(g_{pp\pi_0} - g_{nn\pi_0}) / g_{NN\pi}$ 
we obtain an interval of $1.2 * 10^{-2}$ to $3.7 * 10^{-2}$,
the uncertainties coming mainly from the  input parameters.
The charged pion nucleon coupling is found to be the average of
$g_{pp\pi_0}$ and $g_{nn\pi_0}$. Electromagnetic effects are not included.
\end{abstract}

The effect of isospin violating meson nucleon couplings has recently seen 
a strong revival of interest in the investigation of charge
symmetry breaking (CSB) phenomena \cite{Pie,GHP1,GHP2,KFG} (for a 
comprehensive review see \cite{MNS} and references therein).
On a microscopical level, isospin symmetry is broken by the electromagnetic
interaction as well as the mass difference of up and down quarks
$m_u \ne m_d$.
It is the aim of this paper to examine the difference between the
pion nucleon coupling constants 
$g_{pp\pi^0}$, $g_{nn\pi^0}$ and $g_{pn\pi^+}$ using the QCD sum rule 
method, which has been established as a powerful and fruitful technique for describing hadronic 
phenomena at intermediate energies \cite{SVZ,RRY1,Nar1}.
Here we will only look at effects which arise from isospin breaking 
in the strong interaction. In the QCD sum rule method this is reflected by
$m_u \ne m_d$ as well as by the isospin breaking of the vacuum condensates.
Electromagnetic effects are  not examined.
Our work follows the approach of refs. \cite{RRY1,RRY2,Rei} and extends
their analysis to the isospin violating case.

We start from the three point function of two nucleonic (Ioffe) \cite{Iof} 
and one pseudoscalar interpolating currents with the appropriate isospin
quantum numbers \cite{RRY1,RRY2,Rei,NP,Mei}, e.g.:
\begin{equation}
A_{NN\pi^i} (p_1,p_2,q) = 
\int d^4 x_1 d^4 x_2 e^{ip_1 x_1} e^{- ip_2 x_2}
\left \langle 0 \vert {\cal T} 
\eta_N (x_1) P^{T=1} _i (0) {\bar{\eta_N}} (x_2) \vert 0 \right \rangle ,
\label{eq1}
\end{equation}
where $i$ stands for $+$ or $0$ and $N$ for proton or neutron, respectively.
The expressions for the pseudoscalar isovector currents read
\begin{mathletters}
\label{eq2}
\begin{eqnarray}
P^{T=1} _{i=0} (x) &=& 
{\bar{u}} (x) i \gamma_5  u ( x) -
{\bar{d}} (x) i \gamma_5  d ( x) ,  \\
P^{T=1} _{i=+} (x) &=& \sqrt{2}
{\bar{u}} (x) i \gamma_5  d ( x) ,
\end{eqnarray}
\end{mathletters}
and those for the Ioffe currents are
\begin{mathletters}
\label{eq3}
\begin{eqnarray}
\eta_p (x) &=& \epsilon_{abc} \left [ \left ( u^a (x) {\cal C} \gamma_\mu
u^b (x) \right ) \gamma_5 \gamma^\mu d^c (x) \right ] , \\
\eta_n (x) &=& \epsilon_{abc} \left [ \left ( d^a (x) {\cal C} \gamma_\mu
d^b (x) \right ) \gamma_5 \gamma^\mu u^c (x) \right ] .
\end{eqnarray}
\end{mathletters}
The momenta $p_1$ and $p_2$ are those of the nucleon, and 
$q = p_1 - p_2$ that of the pion; 
${\cal C} = i \gamma_2 \gamma_0$ is the charge conjugation matrix.
In the following we will only keep terms up to first order
in isospin violation, i.e. $m_d - m_u$.

The phenomenological side of the QCD sum rules 
for the three point functions $A$ are obtained by saturating the 
general expressions for the $A$'s (\ref{eq1}) with the 
corresponding nucleon and pion intermediate states.
In order to connect to hadronic observables we have to
know the overlap between the pion states and the interpolating fields.
The axial Ward identity 
\begin{equation}
\partial^\mu {A_\mu}^a = i {\bar q} \gamma_5 \{ {\cal M} , \frac{\tau^a}{2} \} q
\label{eq:awi}
\end{equation}
gives
\begin{mathletters}
\label{eq4}
\begin{eqnarray}
m_0
\langle 0 \vert {\bar u} i \gamma_5 u - {\bar d} i \gamma_5 d 
\vert \pi^0 \rangle 
&=& {m_{\pi^0}}^2 {f_{\pi^0}}\, + \, {\cal O} ((m_u - m_d)^2 ) \\
 \sqrt{2} \, m_0 \, 
\langle 0 \vert {\bar u} i \gamma_5 d 
\vert \pi^+ \rangle
&=& {m_{\pi^+}}^2 {f_{\pi^+}} \, + \, {\cal O} ((m_u - m_d)^2 ) 
\end{eqnarray}
\end{mathletters}
with 
$q = {u \choose d} $,
${\cal M} = m_0 \openone + \frac{m_u - m_d}{2} \tau^3$ and 
$m_0 = \frac{m_u + m_d}{2}$. \\
Hereby we have used that 
$ (m_u - m_d)  \langle 0 \vert {\bar u} i \gamma_5 u + {\bar d} i \gamma_5 d 
\vert \pi^0 \rangle = {\cal O}((m_u - m_d)^2 )$.
Furthermore we can set
${m_{\pi^0}}^2  = {m_{\pi^+}}^2 = {m_{\pi}}^2 $
as well as 
${f_{\pi^0}}  = {f_{\pi^+}} = {f_{\pi}} $,
because the differences between the charged and the neutral quantities are also of 
${\cal O}((m_u - m_d)^2 )$ \cite{GL2}. \\
We also need the current algebra relation
\begin{equation}
m_0 \langle {\bar u} u + {\bar d} d  \rangle 
= (-) {m_\pi}^2 {f_\pi}^2 + {\cal O}((m_u - m_d)^2 )
\label{eq4b}
\end{equation}
which follows from eq.(\ref{eq4})
and the PCAC relation: 
\begin{equation}
\partial^\mu {A_\mu}^a = {m_\pi}^2 f_\pi \pi^a .
\label{pcac}
\end{equation}

The pion nucleon couplings are defined through the interactions:
\begin{mathletters}
\label{eq5}
\begin{eqnarray}
{\cal L}_{pp\pi^0} &=&     {g_{pp\pi^0}} {\bar p} i \gamma_5 \pi^0 p , \\
{\cal L}_{nn\pi^0} &=& (-) {g_{nn\pi^0}} {\bar n} i \gamma_5 \pi^0 n , \\
{\cal L}_{pn\pi^+} &=& \sqrt{2} {g_{pn\pi^+}} {\bar n} i \gamma_5 \pi^+ p .
\end{eqnarray}
\end{mathletters}
It should be remarked that in our notation all three couplings are positive
and have the same value in the isospin conserving limit.

We then obtain the following expressions for the phenomenological sides of
the three point functions, eqs.(\ref{eq1}):
\begin{mathletters}
\label{eq6}
\begin{eqnarray}
A_{pp\pi^0} &=& \, i \, {\lambda_p}^2 \, 
\frac{{m_\pi}^2 f_\pi}{m_0} \,
\frac{(+) g_{pp\pi^0}}{-q^2 + {m_\pi}^2} \,
\frac{1}{{p_1}^2 - {M_p}^2} \,
\frac{1}{{p_2}^2 - {M_p}^2} \,
M_p \, \gamma_5 \, \not{\! q} \,
\,\, + \,\, \dots ,  \label{eq6a} \\
A_{nn\pi^0} &=& \, i \, {\lambda_n}^2 \, 
\frac{{m_\pi}^2 f_\pi}{m_0} \,
\frac{(-) g_{nn\pi^0}}{-q^2 + {m_\pi}^2} \,
\frac{1}{{p_1}^2 - {M_n}^2} \,
\frac{1}{{p_2}^2 - {M_n}^2} \,
M_n \, \gamma_5 \, \not{\! q} \,
\,\, + \,\, \dots ,  \label{eq6b} \\
A_{pn\pi^+} &=& \, i \, {\lambda_p}{\lambda_n} \, 
\frac{{m_\pi}^2 f_\pi}{m_0} \,
\frac{\sqrt{2}g_{pn\pi^+}}{-q^2 + {m_\pi}^2} \, 
\frac{1}{{p_1}^2 - {M_p}^2} \,
\frac{1}{{p_2}^2 - {M_n}^2} \,
\frac{M_p + M_n}{2} \,
\gamma_5 \, \not{\! q} \,
\,\, + \,\, \dots , \label{eq6c}
\end{eqnarray}
\end{mathletters}
where the $\lambda_N$ 's are the overlaps between the Ioffe 
currents (\ref{eq2}) and the corresponding single nucleon states.
The $\dots$ denote contributions from higher resonance intermediate states
and the continuum.
We will come back to these contributions later. 

By saturating the three point function eq.(\ref{eq1}) for the neutral current
with pseudoscalar isovector intermediate states and 
deriving eqs.(\ref{eq6a})
and (\ref{eq6b}) we have assumed so far that the $\pi^0$ mass eigenstate 
is a pure isovector state.
However due to $\pi - \eta$ mixing the correlator in eq.(\ref{eq1}) with
the current $P^{T=1} _{i=0}$
will pick up a contribution from the $|\eta >$
state as well\footnote{We are grateful to K. Maltman
for pointing this out to us.}.
In order to avoid this we have to use a correlator where the pseudoscalar meson
current has only overlap with the physical $| \pi >$, i.e. the mass eigenstate
and not with the $|\eta >$.
As it has been shown in ref. \cite{GL2} this is possible in lowest order chiral
perturbation theory by using 
the linear combination
of the $SU(3)$ flavor octet pseudoscalar currents 
\begin{equation}
P_{a=3}  + \theta P_{a=8}
\label{lin}
\end{equation}
where
\begin{eqnarray}
P_{a=3} &=& {\bar{u}} (x) i \gamma_5  u ( x) - {\bar{d}} (x) i \gamma_5  d ( x) \equiv P^{T=1}_{i=0}
\nonumber \\
P_{a=8} &=& 
\frac{1}{\sqrt{3}} \left [
{\bar{u}} (x) i \gamma_5  u ( x) + {\bar{d}} (x) i \gamma_5  d ( x) - 
\sqrt{2} {\bar{s}} (x) i \gamma_5 s(x) \right ]
\label{su3}
\end{eqnarray}
rather than the pure isovector current in the correlator eq. (\ref{eq1}). 
The $\theta$ denotes the $\pi - \eta$ mixing angle which defines the mass
eigenstates $|\pi >$ and $|\eta >$ in terms of the flavor octet eigenstates
$|\pi_{a=3} >$ and $|\pi_{a=8} > $: 
\begin{mathletters}
\label{eqeps}
\begin{eqnarray}
|\pi >  &= |\pi_{a=3} > + \theta \, |\pi_{a=8} > \label{eqepsa} \\
|\eta>  &= |\pi_{a=8} > - \theta \, |\pi_{a=3} > \label{eqepsb}
\end{eqnarray}
\end{mathletters}
It should be noted in this context that there exists actually a whole family of possible 
choices for interpolating currents involving linear combinations
of $P_{a=8}$ and the flavor singlet current $P_{a=0}$,
which have no overlap with  the $\eta$ but only with the $\pi$.
Our choice (\ref{lin}) is the appropriate one if one ignores
possible mixing to the $SU(3)$ flavor singlet state,
i.e. the $\eta'$, because in this case the current (\ref{lin}) is the only choice which 
has no overlap with the flavor singlet state 
either.

Furthermore it should be noted that 
we have neglected all higher pseudoscalar, isovector 
resonances $\pi^{\prime}$, $\pi^{\prime\prime}$ , $\dots$.
In other words we have assumed that pion pole dominance works at spacelike 
$q^2 \approx - 1 {\mbox GeV}^2$, where the three point function method can be applied
\cite{RRY1,RRY2}. We will discuss this point later as well. 

The next step is to perform the operator product expansion (OPE)
for the three point functions under consideration.
Typical diagrams are shown in Fig.\ref{fig1}.
Following refs.\cite{RRY1,RRY2,Rei} we keep only terms which are proportional
to $\not{\! q} \gamma_5$ and have a $\frac{1}{q^2}$ pole.
We identify the residua of this pole with one on the phenomenological
side, assuming hereby that $\vert q^2 \vert \gg {m_\pi}^2$, so that the 
pion mass can be neglected in eqs.(\ref{eq6}).
Finally we take ${p_1}^2 = {p_2}^2 = - P^2$ in the equation of 
the pole residua and perform a Borel transformation with respect to $P^2$.
It should be noted that the OPE side contains, of course, also terms which
do not have a $\frac{1}{q^2}$ pole.
They will give rise to a form factor, i.e. a $q^2$ dependence of the
pion nucleon couplings \cite{Mei}, which we do not consider in the present
context.

In our case one can easily convince oneself that up to and including
order 4, only the diagrams in Figs.\ref{fig1}(b) and \ref{fig1}(c),
which contain the quark condensates $<{\bar u}u>$ and $<{\bar d }d>$
contribute.
Diagrams containing the gluon condensate $<G^2>$ (Fig.\ref{fig1}(f)) come
in at order 6, because from dimensional arguments they are proportional
to the current quark masses $m_u$ or $m_d$, respectively.
The mixed condensates $<{\bar u} G \cdot \sigma u>$ and
$<{\bar d} G \cdot \sigma d>$ (Fig.\ref{fig1}(g))
are genuinely of two orders higher than the quark condensates.
Four quark condensates (Fig.\ref{fig1}(h)) 
enter already at order 8.
Because reliable values for the isospin breaking of the mixed condensates
and the four quark condensates are missing, 
we prefer to stop the OPE at order 4 and do not take the
higher order power corrections into account.

Applying the prescription described above one can easily derive the 
Borel sum rules for the three point functions of eq. (\ref{eq1}):
\begin{mathletters}
\label{eqsr}
\begin{eqnarray}
(-)
\frac{1}{\pi^2} \,
\left \{ 
\left [ \frac{5}{6} <{\bar u} u > + \frac{1}{6} <{\bar d} d > \right ]
+ \frac{\theta}{\sqrt{3}}
\left [ \frac{5}{6} <{\bar u} u > - \frac{1}{6} <{\bar d} d > \right ]
\right \}  \, = \nonumber \\
{\lambda_p}^2 \, 
\frac{{m_\pi}^2 f_\pi}{m_0} \,
M_p \,
(+) g_{pp\pi^0} \,
\left ( \frac{1}{M^2} \right )^3 \,
e^{- \frac{{M_p}^2}{M^2} }
\label{eq8a} \\
(-)
\frac{1}{\pi^2} \,
\left \{ 
(-) \left [ \frac{5}{6} <{\bar d} d > + \frac{1}{6} <{\bar u} u > \right ]
+ \frac{\theta}{\sqrt{3}}
\left [ \frac{5}{6} <{\bar d} d > - \frac{1}{6} <{\bar u} u > \right ]
\right \}  \, = \nonumber \\
{\lambda_n}^2 \, 
\frac{{m_\pi}^2 f_\pi}{m_0} \,
M_n \,
(-) g_{nn\pi^0} \,
\left ( \frac{1}{M^2} \right )^3 \,
e^{- \frac{{M_n}^2}{M^2} }
\label{eq8b} 
\end{eqnarray}
\end{mathletters}
and
\begin{equation}
(-)
\frac{1}{\pi^2} \,
\left [ \frac{1}{2} <{\bar u} u > + \frac{1}{2} <{\bar d} d > \right ] 
\, = \,
{\lambda_p}{\lambda_n} \, 
\frac{{m_\pi}^2 f_\pi}{m_0} \,
\frac{M_p +M_n}{2} \,
g_{pn\pi^+} \,
\left ( \frac{1}{M^2} \right )^2 \,
\frac{
e^{- \frac{{M_p}^2}{M^2} } - e^{- \frac{{M_n}^2}{M^2} } }
{{M_n}^2 - {M_p}^2 } \, .
\label{eq8c} 
\end{equation}
It should be noted hereby that the strange quark in the current $P_{a=8}$ (eq.(\ref{lin}))
does not contribute in the OPE up to that order which we are taking into account.

Already at this point 
we see 
by taking the difference between eq.(\ref{eq8a}) and eq.(\ref{eq8b})
and comparing with eq.(\ref{eq8c})
that up to first order in isospin breaking
the charged pion nucleon coupling is exactly the arithmetic average
of the two neutral pion nucleon couplings, i.e. we have:
\begin{equation}
g_{pn\pi^+} = \frac{1}{2} [ g_{pp\pi^0} + g_{nn\pi^0} ] .
\label{eq9}
\end{equation}
which is a simple consequence of the $u$ and $d$ quark contents of
the three point functions and valid within the approximations considered.

In order to obtain the splitting between $g_{pp\pi^0}$ and 
$ g_{nn\pi^0}$ we take the sum between eq.(\ref{eq8a}) and
eq.(\ref{eq8b}) and divide by either one of them.
Expanding again up to first order in isospin breaking, we obtain:
\begin{equation}
\left ( - \frac{\delta g}{g_{\pi NN}} \right ) \,
+
\left ( - \frac{\delta {\lambda_N}^2}{{\lambda_N}^2} \right ) \, 
+
\left ( - \frac{\delta M_N}{M_N} \right ) \,
+
2 \left ( \frac{\delta M_N}{M_N} \right )
\left ( \frac{{M_N}^2}{M^2} \right ) \,
= \, - \frac{2}{3} \gamma \, + \, \frac{4}{3} \frac{\theta}{\sqrt{3}} .
\label{eq10}
\end{equation}
Here we have used the following notations
for the isospin splittings:
\begin{equation}
\delta M_N = M_n - M_p , \:
\delta g = g_{nn\pi^0} - g_{pp\pi^0} , \:
\delta {\lambda_N}^2 = {\lambda_n}^2 - {\lambda_p}^2 ,
\label{eq11}
\end{equation}
and the average values
\begin{equation}
M_N = \frac{1}{2} (M_n + M_p) , \:
g_{NN\pi} = \frac{1}{2} (g_{nn\pi^0} + g_{pp\pi^0}) , \:
{\lambda_N}^2 = \frac{1}{2} ({\lambda_n}^2 + {\lambda_p}^2 ) .
\label{eq12}
\end{equation}
Furthermore we have introduced the parameter 
\begin{equation}
\gamma = \frac{<{\bar d}d>}{<{\bar u}u>} - 1 
\label{eq13}
\end{equation}
to denote the isospin breaking in the quark condensates
and set 
\begin{equation}
<{\bar q}q> = \frac{1}{2} [<{\bar u}u> + <{\bar d}d> ] .
\end{equation}

 From eq.(\ref{eq10}) we also see that we need to know the value of
$\delta{\lambda_N}^2$, i.e. the isospin breaking in the overlaps between
the nucleon states and the corresponding interpolating currents.
To obtain $\delta{\lambda_N}^2$, we follow
refs. \cite{RRY1,RRY2,Rei}
and use the sum rules for the 
nucleon two point functions 
\begin{equation}
\int d^4 x e^{ik x} 
\left \langle 0 \vert {\cal T} 
\eta_N (x)  {\bar{\eta_N}} (0) \vert 0 \right \rangle   
= \not{\! k} \Pi^N _1 (k^2) + \Pi^N _2 (k^2) , 
\end{equation}
which have been considered in the case of isospin breaking in refs. 
\cite{HHP1,HHP2,YHHK}.
We will take the chiral odd sum rule for the amplitudes $\Pi_1 (k^2)$ which is known
to work better than the chiral even ones for $\Pi_2 (k^2)$ \cite{HP}.
Including again condensates up to order 4 we have (c.f.eqs.(8) and (11) in ref.\cite{YHHK}):
\begin{mathletters}
\label{eq14}
\begin{eqnarray}
(2 \pi)^4
\frac{{\lambda_p}^2}{4} 
&=&
e^{\frac{{M_p}^2}{M^2} }
\left [ 
\frac{M^6}{8} 
+ \frac{ M^2 g_c ^2 <G^2>}{32} + 
(2 \pi)^2
\frac{M^2}{4}
m_d <{\bar d}d > 
\right ] ,
\label{eq14a} \\
(2 \pi)^4
\frac{{\lambda_n}^2}{4} 
&=&
e^{\frac{{M_n}^2}{M^2} }
\left [ 
\frac{M^6}{8} 
+ \frac{ M^2 g_c ^2 <G^2>}{32} + 
(2 \pi)^2
\frac{M^2}{4}
m_u <{\bar u}u > 
\right ] ,
\label{eq14b} 
\end{eqnarray}
\end{mathletters}
It should be noted that in this sum rule the gluon condensate
$g_c ^2 <G^2>$ enters in the same order as the quark condensate
(i.e. order 4)
and therefore is taken into account, whereas in the sum rule 
for the three point function (eqs.(\ref{eq8a})-(\ref{eq8c})) 
it enters two orders
higher than the quark condensate and was therefore omitted.

We take the difference between eq.(\ref{eq14a}) and  eq.(\ref{eq14b})
for $p$ and $n$
and divide
by either one of them, giving 
\begin{eqnarray}
\left ( - \frac{\delta {\lambda_N}^2}{{\lambda_N}^2} \right ) \,
&=& (-2) \left ( \frac{\delta M_N}{M_N} \right ) 
\left ( \frac{{M_N}^2}{M^2} \right ) \,
\nonumber \\ 
&-& 
(2\pi)^2 ({m_\pi}^2 {f_\pi}^2)
\frac{M^2}{M^6 + \frac{1}{4} g_c ^2 <G^2> M^2}
\left[ 2 \frac{m_d - m_u}{m_d + m_u} + \gamma - 2  \frac{\delta M_N}{M_N} 
 \frac{{M_N}^2}{M^2} \right ]  .
 \label{eqll}
\end{eqnarray}
Putting eq.(\ref{eqll}) into eq.(\ref{eq10}) we obtain the 
final sum rule \begin{eqnarray}
\left ( - \frac{\delta g}{g_{NN\pi}} \right ) \,
&=& - \frac{2}{3} \gamma \, + \frac{4}{3} \frac{\theta}{\sqrt{3}} \, 
+
\left ( \frac{\delta M_N}{M_N} \right ) \nonumber \\
&+&
(2\pi)^2 ({m_\pi}^2 {f_\pi}^2)
\frac{M^2}{M^6 + \frac{1}{4} g_c ^2 <G^2> M^2 }
\left[ 2 \frac{m_d - m_u}{m_d + m_u} + \gamma - 2  \frac{\delta M_N}{M_N} 
 \frac{{M_N}^2}{M^2} \right ] ,
\label{eq15}
\end{eqnarray}
where we have used eq.(\ref{eq4b}).

For the isospin breaking in the quark masses we use the most recent
analysis of current quark mass ratios \cite{Leu}, giving a value of
$\frac{m_d - m_u}{m_d + m_u} = 0.29 \pm 0.05$ \\
As we can see,  
one of the crucial ingredients in eq.(\ref{eq15}) is the numerical value for 
the parameter $\gamma$.
Various analyses concerning this quantity have been performed using different methods:
QCD sum rules for scalar and pseudoscalar mesons
\cite{Nar1,Nar2,DD,Nar3},
QCD sum rule analyses of the the baryon mass splittings \cite{ADJ} and the $D$ and $D^*$ isospin
mass differences \cite{EI} as well as
effective models for QCD incorporating
the dynamical breaking of chiral symmetry \cite{HHP2,PRS}.
The range for $\gamma$ resulting from these analyses is rather large: 
$0.002 < - \gamma < 0.010$. This range is also consistent with 
the result obtained from 1-loop chiral perturbation theory assuming reasonable values for
the strange quark condensate $\langle {\bar s} s \rangle$ \cite{GL2}. 

For the $\pi - \eta$ mixing angle $\theta$ we take the value obtained in lowest
order chiral perturbation theory \cite{GL2}: 
\begin{equation}
\theta = \frac{1}{4} \sqrt{3} \frac{m_d - m_u}{m_s - m_0}
\label{epsdef}
\end{equation}
Using the numerical values for the quark mass ratios from ref. \cite{Leu} we find
$\theta = (10 \pm 0.8) * 10^{-3}$.
Next to leading order corrections are typically of the order $30 \%$, e.g. the decay
constants $f_\pi$ and $f_\eta$ differ by about $30 \%$ if loops are included.
It seems therefore appropriate to assign an error of $30 \%$ to the contribution
coming from $\pi - \eta$ mixing, i.e. to the term $\frac{4}{3} \frac{\theta}{\sqrt{3}}$ in 
eq.(\ref{eq15}).
We have already mentioned that
in the treatment of the $\pi$-$\eta$
mixing we have ignored the mixing between $\eta$ and $\eta^\prime$ as well
as $\pi$ and $\eta^\prime$. The treatment of the $\eta^\prime$ in the current
approach is difficult due to the anomaly in the SU(3) singlet pseudoscalar current.
The value of the $\pi$-$\eta$ mixing angle $\theta$ increases by about $30 \%$,
if the $\eta^\prime$ is included \cite{CS}.

In order to obtain $\delta M_N$, we correct the experimental value for the
proton and neutron mass difference by electromagnetic effects, rendering
an interval of $1.6{\mbox{MeV}} < \delta M_N < 2.4 {\mbox{MeV}}$ 
\cite{GL1,HM}.
For $g_c ^2 <G^2>$ we take the standard value of 
$0.474 {\mbox{GeV}}^4$, noting that its
numerical contribution to eq.(\ref{eq15}) is rather small.

The dashed curve of Fig.\ref{fig2} shows 
$\left ( - \frac{\delta g}{g_{NN\pi}} \right ) $ 
obtained from
eq.(\ref{eq15}) in the Borel window 
$0.7 \mbox{GeV}^2 < M^2 < 1.5 \mbox{GeV}^2$ using typical values
for the parameters.

Up to now we have saturated the phenomenological side of the sum rules
only with the $N$ ground state and have omitted transitions between 
$N$ and excited $N^*$ states as well as contributions from the pure continuum.
As has been shown e.g. in refs. \cite{IS,Iof2,BK}
in a single variable dispersion sum rule,
the transitions $N \to N^*$ gives rise to a single pole term
$ \sim \frac{1}{p^2 - {M_N}^2 }$ in addition to the double pole 
term of eq.(\ref{eq6}). 
This single pole term will not be suppressed in the Borel
sum rules (\ref{eq15}).
It is easy to see that the inclusion of this contribution
would add a term to the l.h.s. of eq.(\ref{eq15}) which is
of the same general form multiplied by an additional power of $M^2$,
i.e. it can be written as 
$C \left ( \frac{1}{M^2} \right )^2 e^{- \frac{M_N ^2}{M^2} }$.
The constant $C$ can be treated as effective parameter which
is optimized in order to obtain the best fit to the Borel curve.
In the isospin conserving case \cite{RRY1,RRY2,Rei} it seems to be
justified to neglect this contribution
due to the fact that the sum rule
for $g_{NN\pi}$ saturated only with the ground state 
is practically independent on the Borel mass $M^2$.
This indicates that the 
parameter $C$ is compatible with zero.
Furthermore the on shell value for $g_{NN\pi}$ is reproduced rather
well in this approach.
A recent QCD sum rule analysis for $g_{NN\pi}$ using 
two point functions \cite{BK} also finds
that this transition is very small.
However, in our case we are looking at isospin violation, and there could
be a small difference of the parameter $C$ for the proton and neutron
contributing to the sum rule (\ref{eq15}) in the same order of magnitude
as $\frac{\delta g}{g_{NN\pi}}$.
It is not difficult to take the excited states into account.
The l.h.s. of eq.(\ref{eq15}) becomes 
$\left ( - \frac{\delta g}{g_{NN\pi}} \right ) + A M^2 $, where the unknown
parameter $A$ is optimized in the Borel analysis, which means, effectively,
by fitting a straight line to the dashed curve of Fig.\ref{fig2}.
Doing so results in the full line curve of Fig.\ref{fig2} as the final
Borel curve for    
$\left ( - \frac{\delta g}{g_{NN\pi}} \right ) $, which is very 
stable in the window under consideration. \\
It should be noted that the $M^2$ dependence of the Borel curve is practically unaffected 
by the large uncertainty in the input parameter $\gamma$ and only depends on the ratio 
$\frac{m_d - m_u}{m_d + m_u}$, because 
the numerical contribution of $\gamma$ as well as $\delta M_N $
to the $M^2$ dependent term in eq.(\ref{eq15})
is very small. 
The term $- \frac{2}{3} \gamma + \frac{4}{3} \frac{\theta}{\sqrt{3}} $ 
only affects the intersection with the $y$-axis but not
the $M^2$ dependence.

Finally let us look at the effect of a pure continuum starting at
a threshold $s$, which would result in multiplying the r.h.s
of the eqs.(\ref{eq8a})-(\ref{eq8c}) 
with the function $E_1 (x) = 1 - (1+x) e^{-x}$ with
$x = \frac{s}{M^2}$.
If one assumes that the continuum thresholds for proton $s_p$ and 
neutron $s_n$ are equal, there is no effect to the isospin breaking
sum rule (\ref{eq15}).
Allowing for a difference of $\frac{|\delta s|}{s} = 0.2 \%$ 
(compatible with $\frac{\delta{M_N}}{M_N}$), with $s_n > s_p$ 
and using a typical
value of $s= 2.25 \mbox{GeV}^2$ would give a contribution
of $\approx 0.17 \% $ to $\left ( - \frac{\delta g}{g_{NN\pi}} \right )$ 
at $M^2 = 1 \mbox{GeV}^2$.
This is noticeably smaller than other errors inherent in the sum 
rule method.
   
In order to obtain an estimated error for
$\left ( - \frac{\delta g}{g_{NN\pi}} \right )$ 
we calculate
the minimum  and the maximum values obtained from eq.
(\ref{eq15}) after fitting the constant $A$ 
and using the extreme values for the input
parameters $- \gamma$, $\frac{m_d - m_u}{m_d + m_u}$, $\delta M_N$ as well
as the the contribution from $\pi - \eta$ mixing, as discussed above.
This gives an interval of
\begin{equation} 
17 * 10^{-3} < \left ( - \frac{\delta g}{g_{NN\pi}} \right ) < 30 * 10^{-3} .
\label{interval1}
\end{equation}
Furthermore, from various other isospin violating sum rule analyses (e.g. ref.\cite{YHHK})
we know that the next higher order condensate 
$\langle {\bar q} G\cdot \sigma q \rangle$, which has
been omitted here due to the reasons mentioned above, may account
for about $25 \%$ of the leading term.
This means that we can expect an additional uncertainty
of this magnitude.
This leaves us with a final interval of
\begin{equation} 
12 * 10^{-3} < \left ( - \frac{\delta g}{g_{NN\pi}} \right ) < 37 * 10^{-3} .
\label{interval2}
\end{equation}
The contribution coming from $\pi$-$\eta$ mixing, the term $\frac{4}{3} \frac{\theta}{\sqrt{3}}$
in eq.(\ref{eq15}) amounts to about $8\pm 2.5 *10^{-3}$.
As stated above this value would be about $30\%$ larger if $\eta - \eta^\prime$ mixing was included. 
The large uncertainty in the input parameter $\gamma$
and the lack of phenomenological data do not call for a 
more detailed investigation at the present stage.

Finally let us compare our result with those of previous studies, which 
analyze the isospin splitting of the pion nucleon couplings arising from 
the strong interaction, i.e.,  essentially
the quark mass difference $m_d -m_u$. 
It should be noted that direct experimental values are not available.
The Nijmegen phase shift analysis for $NN$ and 
$N {\bar N}$ scattering data \cite{TRS,STS}
which is consistent with data from $\pi N$ scattering \cite{ALRW},
but includes electromagnetic effects, 
finds $\left ( \frac{\delta g}{g_{NN\pi}} \right )$ = 0.002, but with an 
error of 0.008; thus, there is no evidence for a difference and they also 
find no evidence for a difference between $g_{pn\pi^+}$ and $g_{NN\pi^0}$
within the statistical errors of their analysis.

 From table \ref{tab1} we see that our range for 
$\left ( - \frac{\delta g}{g_{NN\pi}} \right )$ 
in (\ref{interval2}) is  compatible with the values 
obtained by other authors, both in sign and order of magnitude: \\
(1) The quark gluon model of Henley and Zhang \cite{HZ}; \\
(2) the quark pion model of 
Mitra and Ross \cite{MR,MNS}. This has recently been used by
Piekarewicz \cite{Pie}, who obtained a violation of the ``triangle identity''
consistent with the $\pi N$ data analysis of ref.\cite{GAK}; \\
(3) the use of the quark mass difference $m_d - m_u$ and $\pi - \eta$ mixing  
\cite{AR};  \\
(4) the chiral bag model \cite{CH}, which also has 
our relation (eq.(\ref{eq9})) for
the charged coupling  to be valid. \\
(5) On the other hand, the use of the cloudy bag model \cite{TBG} leads to 
$\left ( \frac{\delta g}{g_{NN\pi}} \right ) \approx 0.006 $, with the 
opposite sign to our result. 

It should be noted that there are electromagnetic corrections,
whose direction are unknown.
The charge difference we obtain due to the strong interaction would, by 
itself, lead to a difference on the scattering lengths $\vert a_{nn} \vert - 
\vert a_{pp} \vert \approx -0.5 \pm 0.2 {\mbox{fm}} $, smaller than, but 
in the opposite direction to the observed difference \cite {MNS}. 
Of course, there are other effects which play a role, 
e.g. $\rho-\omega$ mixing.

We are aware that using the three-point function is a priori less
suitable than the two-point function for calculating the pion nucleon 
coupling on shell,
because it works at spacelike $q^2 \approx - 1{\mbox GeV}^2$
and needs the detour of comparing the 
$\frac{1}{q^2}$ pole residua \cite{RRY1,RRY2,Rei}.
As we have already mentioned earlier this means that one has to assume that
$\pi$ pole dominance 
can still be applied in this region and higher 
pseudoscalar resonances are neglected, or, in other words
we use the PCAC interpolating pseudoscalar field at those values of $q^2$.
A quantitative analysis of the contribution of these higher resonances would require some knowledge
about their coupling to the nucleon. 
There are various indications and consistency checks that the concept 
proposed in refs. \cite{RRY1,RRY2,Rei} is a reasonable one:
First of all 
in the isospin violating case one is likely to be 
less sensitive to the higher resonance contributions due to cancellation which presumably occur
if summing over the higher pionic excitations.
Furthermore 
in the isospin conserving case \cite{RRY1,RRY2,Rei}
the experimental value of $g_{NN\pi}$ is reproduced rather well.
The Borel stability in both the 
isospin conserving
\cite{RRY1,RRY2,Rei} and the isospin violating case (this work)
is a further consistency check, although this of course only a necessary but not
sufficient condition.
Finally there is the analysis of the $\pi NN$ formfactor \cite{Mei} using
this approach. 
The $q^2$ dependence of $g_{NN\pi} (q^2)$, which contains effectively 
the higher pionic resonances in the spectral function  
(\ref{eq1}), 
is consistent with various other approaches using the same interpolating current but
working at lower $q^2$.
This indicates that the interpolation between low and high $q^2$ region is done reasonably well.
Despite all these arguments the importance of the higher pseudoscalar 
resonances remains a matter to be settled and
needs further quantitative investigation \cite{maltman}.

On the other hand the use of a two-point function \cite{RRY1,Rei,SH,BK} has other, 
and we believe worse problems in the consideration of isospin violation.
The single nucleon pole sum rule, as it has been used in 
refs.\cite{RRY1,Rei,SH}, suffers principally from the 
problem that the contribution from the transition $N \to N^*$ 
enters exactly in the same form 
as 
$g_{NN\pi}$ itself, namely as single pole.
After Borel transform
one obtains a term $g_{NN\pi} + A$ instead of 
$g_{NN\pi} + A M^2$ as in case of the double pole sum rule.
Hence within the single pole sum rule itself there is a priori
no way to separate the $N\to N^*$ contribution $A$ from $g_{NN\pi}$.
In refs.\cite{RRY1,Rei,SH} the $N\to N^*$ transition has been ignored.
In the isospin conserving case 
this is a posteriori justified because
the numerical value of this term turns out to be small,
as it has been discussed above.
However we do not know if this is true in the isospin violating case.
The double nucleon pole sum rule \cite{BK} would avoid this problem
and moreover is able to give a value for the $N\to N^*$ contribution.
Unfortunately as one can see from the analysis in ref.\cite{BK}
already in the isospin conserving case
this sum rule seems to be rather sensitive to the condensate input,
which is the quark condensate $< \bar{q} q>$ 
and especially the higher order mixed condensate 
$\langle 0 \vert {\bar q} {\tilde G} \gamma_{\mu} q \vert \pi \rangle$,
which has to be included to give a reasonable value for $g_{NN\pi}$.
In case of the three point function the condensate input parameters are much better under control.
For these reasons we prefer to work with the three point function sum rule.

To summarize we have calculated the splitting between 
the pion nucleon coupling constants
$g_{pp\pi^0}$, $g_{nn\pi^0}$ and $g_{pn\pi^+}$ due to isospin 
breaking in the strong
interaction by using the QCD sum rules for the corresponding 
pion nucleon three point
functions. We have taken OPE diagrams up to order 4 into account.
Our result for the splitting in the neutral couplings is
$ 1.2 * 10^{-2} < \frac{g_{pp\pi^0} - g_{nn\pi^0}}{g_{NN\pi}} < 3.7 * 10^{-2}$.
The charged coupling $g_{pn\pi^+}$ is found to be the average 
of the two neutral ones.

\acknowledgements
This work has been supported by the DOE (grant \# DE-FG06-88 ER 40427)
and the NSF (grant \# PHYS-9310124 and \# PHYS-9319641).
T.M. would like to thank 
the Institute for Nuclear Theory at the University of Washington
in Seattle for its hospitality during various visits while some of 
this work was being carried out. We also thank Kim Maltman, Jerry Miller 
and Bira van Kolck for helpful comments.

\begin{figure}
\caption{Diagrams in the OPE.}
\label{fig1}
\end{figure}

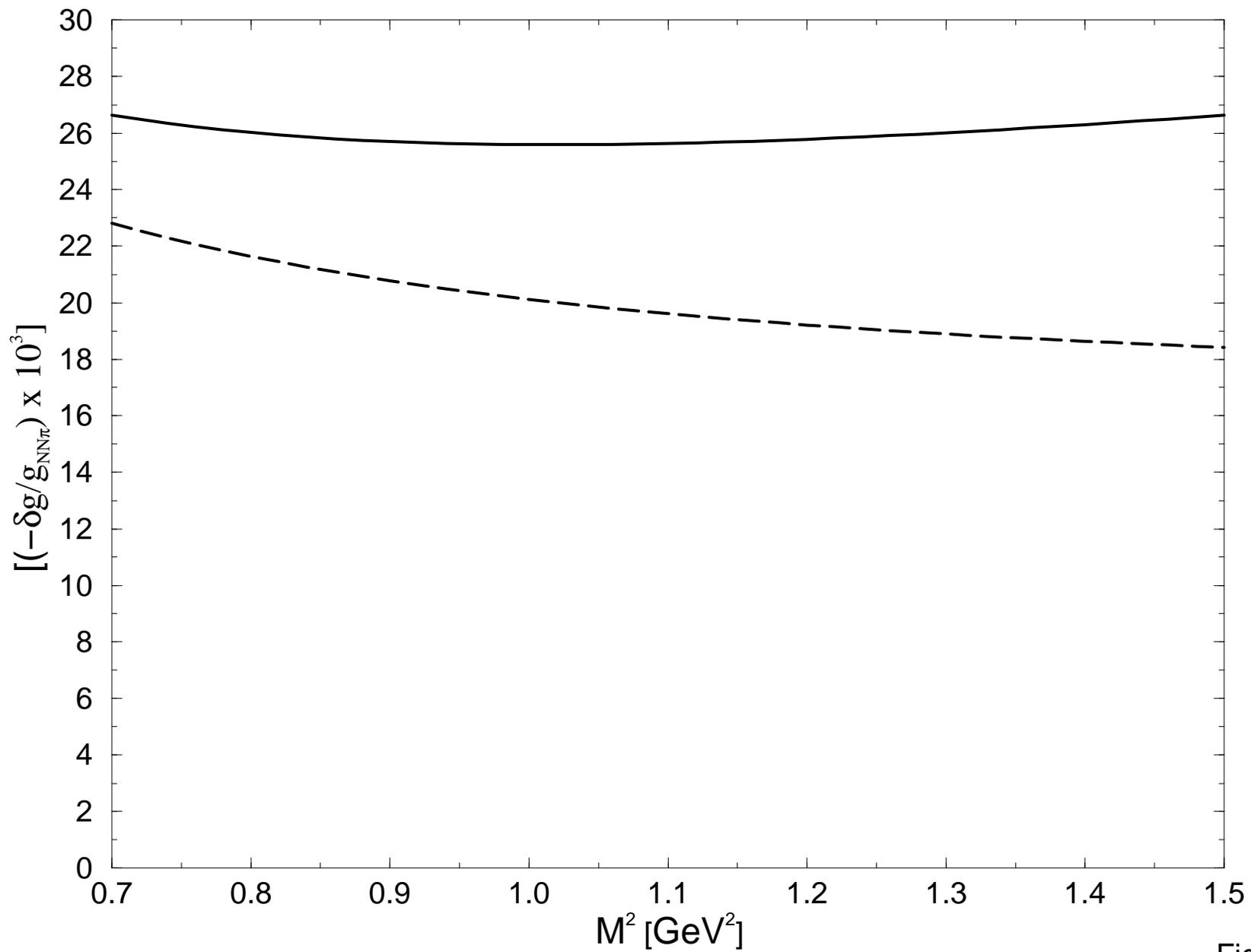
\begin{figure}
\caption{Dependence of 
$({g_{pp{\pi_0}} - g_{nn{\pi_0}}}) / {g_{NN\pi}}$ 
on the square of the Borel 
mass $M^2$.
As example we have used the parameters $\gamma = - 0.01$, 
$\delta {M_N} = 2{\mbox{ MeV}}$ and 
$\frac{m_d - m_u}{m_d + m_u} = 0.28$.
The dashed curve is obtained by omitting the transitions 
$N \to N^*$. 
In the full curve these contributions are included.} 
\label{fig2}
\end{figure}

\begin{table}
\caption{Comparison of $({g_{pp{\pi_0}} - g_{nn{\pi_0}}}) / {g_{NN\pi}}$
obtained in different approaches including isospin violation effects from strong
interaction. In order to compare the numerical values of refs.\protect\cite{Pie,HZ} 
with the other results we have used ${\delta{M_N}} / {M_N} = 0.002$.}
\begin{tabular}{cc}  &$({g_{pp{\pi_0}} - g_{nn{\pi_0}}}) / {g_{NN\pi}}$  \\ \tableline
this work & $\approx 0.012 \dots 0.037$ \\
Ref.\protect\cite{HZ} & $\approx 0.010 \dots 0.014$ \\ 
Ref.\protect\cite{Pie} & $\approx 0.006$  \\
Ref.\protect\cite{AR} & $ 0.005 \pm 0.0018$ \\ 
Ref.\protect\cite{CH} & $ \approx 0.0067$ \\
\end{tabular}
\label{tab1}
\end{table}

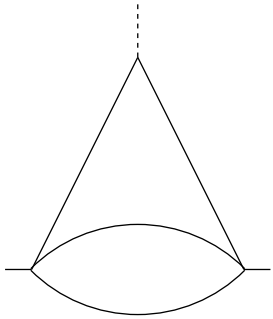
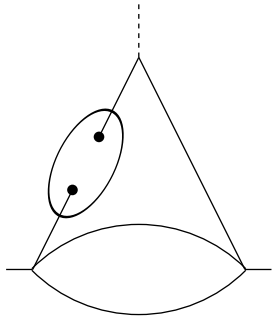
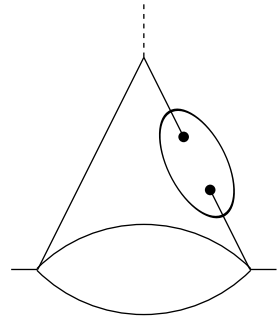
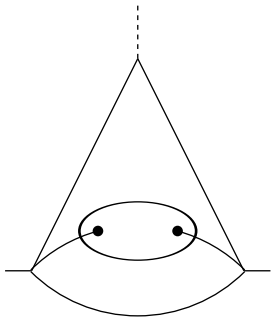
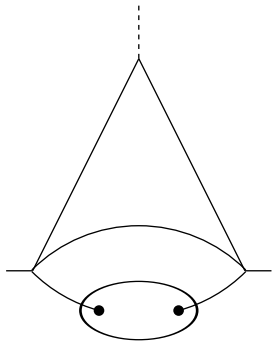
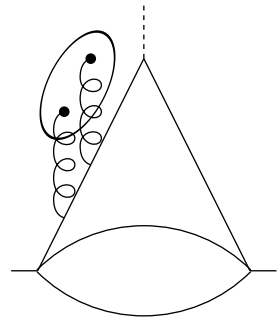
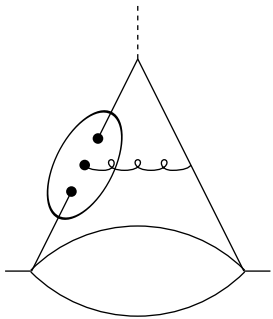
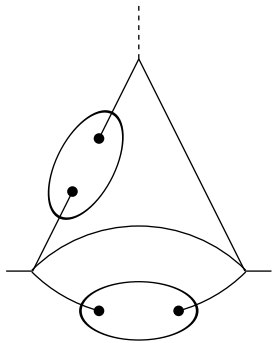
\end{document}